# Emerging Advancements in 6G NTN Radio Access Technologies: An Overview

Husnain Shahid*, Carla Amatetti††, Riccardo Campana††, Sorya Tong**, Dorin Panaitopol**, Alessandro Vanelli-Coralli††, Abdelhamed Mohamed§, Chao Zhang‡, Ebraam Khalifa¶, Eduardo Medeiros‡, Estefania Recayte†, Fatemeh Ghasemifard‡, Ji Lianghai¶, Juan Bucheli¶, Karthik Anantha Swamy¶, Marius Caus*, Mehmet Gurelli¶, Miguel A. Vazquez*, Musbah Shaat*, Nathan Borios∥, Per-Erik Eriksson‡, Sebastian Euler‡, Zheng Li§, Xiaotian Fu‡

* CTTC Spain, † DLR Germany, ‡ Ericsson, § Orange France, ¶ Qualcomm, ∥ Thales Alenia Space France
** Thales SIX France, †† Unibo Italy

*Abstract*—The efforts on the development, standardization and improvements to communication systems towards 5G Advanced and 6G are on track to provide benefits such as an unprecedented level of connectivity and performance, enabling a diverse range of vertical services. The full integration of non-terrestrial components into 6G plays a pivotal role in realizing this paradigm shift towards ubiquitous communication and global coverage. However, this integration into 6G brings forth a set of its own challenges, particularly in Radio Access Technologies (RATs). To this end, this paper comprehensively discusses those challenges at different levels of RATs and proposes the corresponding potential emerging advancements in the realm of 6G NTN. In particular, the focus is on advancing the prospective aspects of Radio Resource Management (RRM), spectral coexistence in terrestrial and non-terrestrial components and flexible waveform design solutions to combat the impediments. This discussion with a specific focus on emerging advancements in 6G NTN RATs is critical for shaping the next generation networks and potentially relevant in contributing the part in standardization in forthcoming releases.

## I. INTRODUCTION

The integration of multilayered Non-Terrestrial Networks (NTN) into 6G (6G NTN) is anticipated to play a pivotal role in enabling global coverage and connectivity, supporting various use cases. The possibility of integrated satellite technology was first studied by 3GPP in Release 15 and 16. Subsequent specification of enhancements to the 5G New Radio (NR) framework in Release 17 aimed at providing basic functionality for the support of 5G via NTN, with the goal to make the minimal but strictly necessary changes to the 5G NR protocol stack to be employed in a satellite system. The specifications are set to undergo further enhancements in the recently concluded Release 18 and the forthcoming Release 19, where a new Work Item [RP-234078] has been agreed. Additionally, this integration is expected to provide the opportunity of overcoming the coverage area limitations and geographical constraints to guarantee ubiquitous connectivity even in remote areas and during the time of natural disasters [1]. The 6G NTN multilayered concept is depicted in Fig 1.

Although NTN integration offers various advantages, it also entails several challenges to optimize the Key Performance Indicators (KPIs), especially related to RATs. One example is spectral efficiency, given the finite available spectrum for satellite communication while the demand for bandwidth is on the rise with the increasing number of users and the evolution of the services. Efficient use of the available spectrum is challenging and requires further optimization to ensure maximum data throughput [2]. Additionally, adaptability to heterogeneous networks and channel conditions considering the non-uniform user load in spatio-temporal scale and the ability to operate in a range of frequency bands are critical.

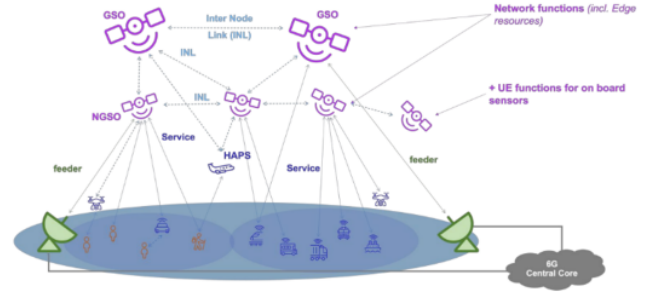

Fig. 1. 3D-multilayered NTN Architecture

In the perspective of RRM, another challenge lies in efficient utilization of limited resources to meet the dynamic throughput needs of users and ensuring fair distribution of resources since unfairness leads to Quality of Service (QoS) disparities. In particular, efficient resource allocation in a multi-orbital and multi-satellite system is critical due to the dynamic nature of inter and intra satellite links and the heterogeneous user traffic demand. In another view, multiple communication systems operating simultaneously within the same frequencies are prone to interference [3].

All these challenges are general in the view of wireless communication systems but critical in 6G NTN RATs due to their unique characteristics and, being confined within the scope of standardization, thereby require substantial study to combat these challenges. In this context, this paper proposes emerging advancements in three different aspects of 6G NTN RATs which directly or indirectly exhibit the potential to combat the aforementioned challenges. Moreover, this paper explicitly instigates an overview of potential solutions within different layers and for various network functions of integrated TN-NTN RATs. A part of the proposed contributions is data-driven AI-based solutions especially for RRM tasks. AI has promising potential towards the optimization of integrated networks due to its learning capabilities, its adaptability to tackle multi-dimensional complex optimization problems.

Considering the significance of the topic, this paper is organized as follows. Section II focuses on the potential perspectives within the framework of flexible waveform for 6G NTN and its applicability to TN system as well, taking into account the robustness towards the intensive impairments built-in present in NTN. Section III is about the profound study of optimizing the RRM with AI and identifying the network functions that are partially explored or yet to be explored among resource allocation solutions in integrated networks. The study of spectrum coexistence in the realm of terrestrial and non-terrestrial networks is comprehensively discussed in Section IV. Finally, the conclusion is drawn in Section V.

## II. FLEXIBLE WAVEFORM ASPECTS IN 6G NTN

This section discusses challenges and research approaches for selecting an improved and converged waveform for 6G, considering both terrestrial and non-terrestrial use cases. We review the characteristics of the waveform in 5G and present some potential candidate solutions for 6G, focusing on the NTN-specific aspects.

The physical layer of 5G NR utilizes Cyclic Prefix Orthogonal Frequency Division Multiplexing (CP-OFDM) and Discrete Fourier Transform-spread OFDM (DFT-s-OFDM) as waveforms. The utilization of OFDM-based transmission brings forth numerous advantages, including resilience to time-invariant frequency-selective channels, high spectral efficiency, and flexibility in terms of slot duration and sub-carrier spacing due to the support for different numerologies. Additionally, OFDM proves highly scalable, enabling the incorporation of additional features to address a wide range of scenarios and applications pertinent to 5G and beyond [4]. In the integration process of TN-NTN, CP-OFDM and DFT-s-OFDM waveforms are adopted as core waveforms of NR NTN. Release 17, 18, and the upcoming Release 19 envision the integration of TN and NTN while proposing minor adjustments to the standard to support NTN. The forthcoming generation of communication networks, namely 6G, is anticipated to exhibit higher data rates, reduced latency, and extreme coverage extension. To fulfill these requirements, high-frequency bands such as centimeter wave, millimeter-wave (mmWave), and TeraHertz (THz) frequencies [5] will play a pivotal role. Additionally, it is widely acknowledged that 6G systems, facilitated by NTN, will harness a fully unified multi-dimensional and multi-layer network where terrestrial and non-terrestrial infrastructures are unified [6]. This unified infrastructure presents numerous challenges, one of which involves defining a unified air interface optimized for both terrestrial and non-terrestrial conditions. In particular, the optimization of the air interface should also involve the efficient design of the waveform in the context of NTN. DL coverage enhancements for smartphones to support NTN are currently within the scope of Release 19 [RP-234078]. These enhancements primarily target outdoor communications between the NTN node and the User Equipment (UE). Consequently, within the realm of 6G, it would be advantageous to delve into more challenging scenarios that demand exploration and evaluation. A list of the desirable air-interface features is provided in Table I.

TABLE I
FEATURES INTRODUCED BY 6G NTN

| Features | Definition |
| --- | --- |
| Compatibility with terrestrial network | If TN and NTN 6G components have a compatible waveform, that would improve 1) seamless handover between the networks, 2) device compatibility, 3) potential to integrate and use 5G/6G infrastructure, 4) coexistence, 5) hardware complexity. |
| Possibility for spectrum sharing | A flexible waveform design enables spectrum sharing between TN and NTN considering different scenarios and use cases. |
| Robustness to co-channel interference | The network should be able to maintain reliable and effective communication in the presence of interference from signals transmitted on the same frequency channels. |
| Possibility for seamless connectivity | The air interface should support seamless connectivity to NTN platforms including flying nodes, NGSO and GSO satellites. |
| Ability to provide an additional link margin in FR1 band | To serve smartphone users in light indoor and in-vehicle conditions, the air interface should be able to provide additional link margin to combat extra penetration loss. |
| Support of accurate network-based positioning | Dedicated pilot signals or reference symbols to facilitate accurate timing and phase measurements for accurate positioning. |
| Support to UE without GNSS | UE should still be able to connect to NTN when GNSS is not available or unreliable, such as indoor environment. |
| Backward compatibility with 5G | An air interface having backward compatibility with 5G will benefit from its flexibility and immediate compatibility with 5G smartphones. |
| Support of FDD and TDD | NTN will operate in both lower and higher bands. Using both FDD and TDD for NTN means that carriers can use multiple frequencies at the same time. The choice of FDD or TDD is scenarios and use cases dependent. |
| Computational complexity | The air interface design should ensure that the network operates effectively while minimizing the demand on processing resources. |

### A. PAPR Reduction Methods

The link between the satellites and the handheld device on ground is characterized by high path losses, due to the long propagation distance and the antenna loss in the device, resulting in low SNR on both UL and DL directions. The most obvious solution would be to increase the Effective Isotropic Radiated Power (EIRP) either in the device or in the satellite, but both approaches are not without challenges. The former is under investigation to be included in the Release 19 scope. The latter might be hard to realize due to the limited power of the satellite, the antenna size limitation on the satellite, and the current limitations of CP-OFDM system which may require additional power backoff. As a matter of fact, the substantial Peak-to-Average Power Ratio (PAPR) in CP-OFDM diminishes the efficacy of the high-power amplifier in the satellite, leading to decreased spectral efficiency. Various techniques have been proposed to mitigate PAPR in OFDM systems, including clipping, coding, non-linear companding, tone reservation and tone injection, SeLective Mapping (SLM), and Partial Transmit Sequence (PTS). Among these methods, the PTS technique is identified as the most efficient and distortion-free approach for reducing PAPR in OFDM systems [7]. Consequently, one avenue for future research could involve an investigation into the application of the PTS method for PAPR reduction in the context of

NTN. Furthermore, in UL, DFT-s-OFDM is widely adopted in 4G and 5G systems, which is characterized by a reduced PAPR. A prospective research direction involves investigating the suitability of DFT-s-OFDM or its variants in the DL as well [8].

### B. OFDM with Reduced OOB

High Out-Of-Band (OOB) emissions may pose challenges for coexistence and spectrum sharing of terrestrial networks with satellite non-terrestrial networks and other systems. Attempting spectrum sharing across unsynchronized systems could result in significant interference, leading to diminished spectrum efficiency. Consequently, appropriate filtering and spectrum shaping design is required. Furthermore, the choice of OFDM variants with reduced OOB emissions may emerge as a viable solution [9], [10] at the expense of increased implementation complexity and high-power consumption [11].

### C. Delay-Doppler domain representation for communications in LoS conditions

In addition to the typical losses of the satellite channel, it is imperative to consider both the characteristics of the propagation environment and those of the satellite constellation. Satellite communications predominantly occur in LoS conditions, which motivates choosing a reduced cyclic prefix length to increase the system throughput. Furthermore, the high speed of the satellite introduces severe Doppler effects, which makes the uplink and downlink synchronization challenging. Expanding on constellation aspects, large satellite constellations often permit the visibility of multiple satellites concurrently. In such scenarios, multi-satellite diversity is leveraged to ensure uniform throughput and enhance system reliability against unexpected obstructions. In this framework, Orthogonal Time Frequency Space (OTFS) and its variations emerge as favorable options because of their ability to operate efficiently for high-speed communications with significant Doppler shifts in rural scenarios where the propagation channel typically includes the LoS component along with few scatterers. Additionally, OTFS achieves a superior communication rate through a per block guard interval, as opposed to a per-symbol cyclic prefix as observed in OFDM. This of course comes at the cost of a more complex channel estimation scheme working on large block-wise operations [12]. In addition, OTFS is characterized by large OOB emission when rectangular pulses are used. When OTFS is implemented using an OFDM modulator/demodulator, OTFS exhibits similar power spectral density and envelope fluctuation as OFDM as depicted in Fig 2. In the simulation, a simple limiter model is used to characterize the power amplifier (PA). A more exhaustive comparative assessment is deferred to future work.

### D. Link Margin Enhancement in NTN

To enhance the link margin in FR1 for extended coverage within light indoor environments (e.g., inside a car), one possible approach is to consider additional coding schemes. The channel coding of 5G NR has been optimized for terrestrial systems, with potential compensation of error rates through adaptive hybrid automatic re-transmission without significantly affecting service latency. However, in the context of NTN, the adoption of such re-transmission schemes would adversely impact the transmission and reception buffers and therefore impact the service latency. Consequently, it may be advantageous to consider schemes providing additional coding gain to exhibit a very low error rate under low SNR conditions.

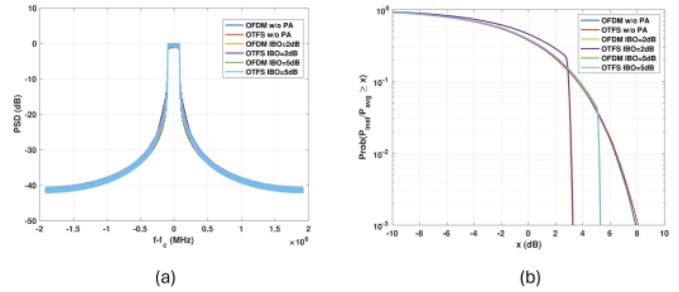

Fig. 2. a) PSD of OTFS and OFDM, (b) PAPR CDF of OFDM vs OTFS. Without and with the HPA with two Input Back-Off values.

## III. RADIO RESOURCE MANAGEMENT ASPECTS

Efficient RRM has paramount significance in integrated TN-NTN networks to ensure QoS requirements and overall system optimization by dynamically allocating the available resources considering the non-uniform user demand and heterogeneous network conditions. This can be done by intelligent decision-making for available throughput capacity, spectrum usage, and mitigating interference to enhance overall network performance. To this end, this section provides AI-based network functions advancements in NTN aspects that are partially explored or yet to be explored in RRM.

### A. Cell Load Balancing

TN deployed in rural areas experience dynamic user traffic throughout the day, resulting in high load for base stations during peak periods and under-utilization during off-peak periods. With the NTN, the load can be balanced between the TN and NTN cells, off-loading part of the terrestrial base stations' traffic. This technique could increased QoS during high network load conditions and a reduced TN energy consumption when the traffic request is lower.

In case of overloaded TN, an AI component can gather information on the network's status and use it to optimize the traffic to be handed over to NTN. The AI's goal is to maximize the overall SINR of TN while accounting for the capacity limitation and high dynamicity of NTN. AI must be able to predict future user traffic and the capabilities of the NTN to ensure that only users with traffic requests compatible with the NTN's performance are handed over. However, there may also be periods of low user traffic, particularly at night, during which TN base stations are kept operational, consuming energy to serve only sporadic users. By utilizing the NTN component, a small amount of traffic could be offloaded from the TN, allowing some of the terrestrial base stations to be turned off and save energy. However, due to the limited throughput and high latency of NTN, not all users are suitable for offloading from TN. To address this issue, an AI component can minimize TN energy consumption while predicting future user traffic and ensuring that users who cannot be adequately served by NTN are not handed over to it.

*B. Fractional Frequency Reuse*

Fractional Frequency Reuse (FFR) is a well-known interference mitigation technique in terrestrial cellular networks but has merely been investigated in multi-beam satellite systems, even though the interference in NTN is more concerning than terrestrial networks and thereby significantly affects the throughput capacity. In essence, the idea is to map and explore the traditional FFR concept into a multi-beam satellite system.

In the process of optimizing the radio resources, adaptive beamforming techniques are preferred to project the flexible beams based on the prior traffic density information and throughput demand. This improves two major factors: 1) the optimization of overall system capacity utilization and 2) fairness in the resource distribution to users in each beam as proven in [13]. However, optimization in this way comes up with the beams overlapping which causes interference and impact the system capacity. The solution to this problem is an investigation of multi-beam FFR schemes and related frequency allocation to avoid interference impairment despite beam overlapping. In essence, the problem boils down to solving the coloring node problem in graph theory where heuristic algorithms have been used to optimize the solution but with more complexities [3]. Herein, the proposal is to employ or redesign multi-agent off-policy Reinforcement Learning (RL) based optimization algorithms for FFR in multi-beam satellite systems.

*C. Traffic Prediction*

Anticipating high volume wireless traffic is crucial not only for overseeing high dynamic and low latency communication networks, but also for improving the quality of provided services, providing a long-term network planning, resource management and optimization, particularly within the context of 6G wireless networks. The prediction process is mainly revolving around exploring and analyzing historical network traffic patterns for assessing future traffic volume and trends. Full system automation and optimization with an iterative system design principle is one of the main 6G network design pillars. Considering the full integration between TN and NTN and the latency constraints for many potential services, the future traffic volume and trends prediction in different sub-networks or geographical areas will facilitate the proactive planning for scaling the network infrastructure and resources accurately, eco-friendly and in a real-time fashion.

*D. Link Quality Prediction*

A distinguishing characteristic of NTN is the fast movement of satellites, at least for LEO constellations, which necessitates frequent satellite-switching by a UE, regardless of whether it is moving or not. To serve the UE continuously during the satellite-switch, the UE needs to be handed over by the network to the next satellite. While there are methods based on timing and UE location, in the baseline solution in 5G NTN, the UE is configured by the network to perform radio measurements of its surrounding satellites and to report the same to the network. The report is then used by the network to understand the UE's radio conditions and to determine the best/next serving satellite for the UE. However, the described online radio measurement procedure in the legacy solution may cause certain inefficiency and bottlenecks, such as power consumption, spectrum usage, potential service interruptions due to measurement gaps, early coordination at network side. Thus, to improve the efficiency during NTN mobility, an AI-based mechanism can be considered and leveraged to estimate and predict the UE's radio channel condition in the time-and-spatial domain, e.g. the radio channels towards different satellites and at different time instances, thereby reducing the dependence on reference signals for measurements.

## IV. SPECTRUM COEXISTENCE ASPECTS

In Release 17, 3GPP considered for the first time the introduction of Mobile Satellite Service (MSS) frequency bands for direct connectivity with satellites and had to consider the coexistence in adjacent bands with TNs. To this end, 3GPP studied the capability of reuniting two different types of services by reusing the same waveform and potentially the same type of terminal. The frequency bands currently defined as part of Release 17, 18 are described in Tables II and III.

TABLE II
NTN OPERATING BANDS IN FR1 FOR SATELLITE NETWORKS (FR1-NTN)

| NTN satellite operating band[a] | UpLink (UL) operating band SAN receive / UE transmit $F_{UL,low} - F_{UL,high}$ | DownLink (DL) operating band SAN transmit / UE receive $F_{DL,low} - F_{DL,high}$ | Duplex mode |
|---|---|---|---|
| n256 | $1980 - 2010$ MHz | $2170 - 2200$ MHz | FDD |
| n255 | $1626.5 - 1660.5$ MHz | $1525 - 1559$ MHz | FDD |
| n254 | $1610 - 1626.5$ MHz | $2483.5 - 2500$ MHz | FDD |

[a] NOTE: FR1-NTN satellite bands are numbered in descending order from n256.

TABLE III
NTN OPERATING BANDS IN ABOVE 10 GHZ FOR SATELLITE NETWORKS (FR2-NTN)

| NTN satellite operating band[a] | UpLink (UL) operating band SAN receive / UE transmit $F_{UL,low} - F_{UL,high}$ | DownLink (DL) operating band SAN transmit / UE receive $F_{DL,low} - F_{DL,high}$ | Duplex mode |
|---|---|---|---|
| n512[1] | $27.5 - 30.0$ GHz | $17.3 - 20.2$ GHz | FDD |
| n511[2] | $28.35 - 30.0$ GHz | $17.3 - 20.2$ GHz | FDD |
| n510[3] | $27.5 - 28.35$ GHz | $17.3 - 20.2$ GHz | FDD |

[1] This band is applicable in the countries subject to CEPT ECC Decision(05)01 and ECC Decision (13)01.
[2] This band is applicable in the USA subject to FCC 47 CFR part 25.
[3] This band is applicable for Earth Station operations in the USA subject to FCC 47 CFR part 25. FCC rules currently do not include ESIM operations in this band (47 CFR 25.202).
[a] NOTE: FR2-NTN satellite bands are numbered in descending order from n512.

FR1-NTN refers to the frequency range 410 MHz – 7125 MHz, and FR2-NTN currently refers to 17300 MHz – 30000 MHz. In Release 17, 3GPP started introducing frequency bands for NTN with bands n256 and n255 (S- and L-bands), operating in FDD mode. Release 18 introduced another band n254 in the FR1-NTN frequency range and added the satellite

Ka-band above 10 GHz with bands n512, n511, and n510. For Release 19, introduction of the Ku satellite band (10 GHz - 14 GHz) is proposed.

One of the major conclusions of the work in Release 17 for FR1-NTN was that an NTN UE could reuse the current RRM and Radio Frequency (RF) requirements of the TN UE. For this reason, at least in FR1-NTN, the same UE (and not only dedicated satellite terminals) can connect to both TN and NTN. The NTN work in 3GPP showed that satellite connectivity does not require a dedicated satellite waveform since 5G New Radio technology based on CP-OFDM (for DL) and DFT-s-OFDM (for UL) can be sufficient for satellite communications. Investigating optimizations to the waveform design, as described in Section II, will also include more realistic NTN-TN coexistence scenarios in adjacent bands by taking into account spectral properties of more advanced waveforms.

### A. Spectrum management scenarios in 6G NTN

Following a similar process as the one proposed by 3GPP (see e.g. TR 38.863 [14]), one of the goals could be to introduce new frequency bands for 6G communications such as NTN Q/V-band (DL: 37.5 – 42.5 GHz (Q-band), UL: 47.2 – 50.2 GHz and 50.4 – 51.4 GHz (V-band)) taking into account current allocations to TN mobile services (DL & UL: 37.0 – 43.5 GHz, DL/UL in some countries (e.g., Brazil): 45.5 – 47 GHz and 47.2 – 48.2 GHz). The results from the NTN-TN adjacent coexistence studies in Q/V band could then be used to derive respective Satellite Access Node (SAN) and Very Small Aperture Terminal (VSAT) UE RF core requirements such as Adjacent Channel Leakage Ratio (ACLR) and Adjacent Channel Selectivity (ACS). In this scenario, Table IV regroups the different interference scenarios identified for Q/V band.

TABLE IV
INTERFERENCE SCENARIOS FOR Q/V BAND

| Associated figure and frequency | N° | Aggressor | Victim | Scope of coexistence simulation |
|---|---|---|---|---|
| Fig. 3 with carrier frequency $f_c = 47$ GHz for simulation purposes | i1 | NTN UL | TN UL | ACLR NTN UE to be varied/defined ACS TN gNB fixed |
| | i2 | TN UL | NTN UL | ACLR TN UE fixed ACS NTN SAN to be varied/defined |
| | i3 | NTN UL | TN DL | ACLR NTN UE to be varied/defined ACS TN UE fixed |
| | i4 | TN DL | NTN UL | ACLR TN gNB fixed ACS NTN SAN to be varied/defined |
| Fig. 4 with carrier frequency $f_c = 37$ GHz for simulation purposes | i5 | TN DL | NTN DL | ACLR TN gNB fixed ACS NTN UE to be varied/defined |
| | i6 | NTN DL | TN DL | ACLR NTN SAN to be varied/defined ACS TN UE fixed |
| | i7 | NTN DL | TN UL | ACLR NTN SAN to be varied/defined ACS TN gNB fixed |
| | i8 | TN UL | NTN DL | ACLR TN UE fixed ACS NTN UE to be varied/defined |

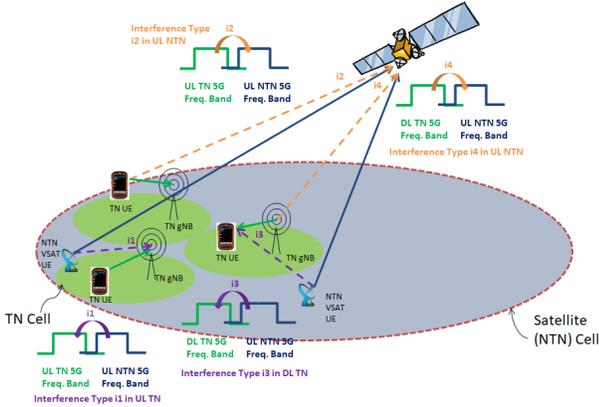

Fig. 3. Exemplification of interference type i1 to i4

The design principles and simulation assumptions are based on 3GPP methodology already used from S-band (Release 17) and Ka-band (Release 18). The simulation assumptions were adapted for Q/V bands, while taking into account physical limitations of VSAT UE, e.g. an aperture of maximum 20x20cm useful for the automotive industry or different other use cases such as aerial communication for drones.

The NTN-TN adjacent band coexistence principle is to derive ACLR for the transmitter and ACS for the receiver by considering different combinations of NTN/TN transmitter-receiver links in both DL and UL. Using 3GPP specifications TS 38.104 [15] (TN BS FR1 and FR2), TS 38.101-2 [16] (TN UE FR2), and TS 38.101-1 [17] (TN UE FR1), derived for the terrestrial networks, the considered assumptions for the TN ACLR and ACS values are summed up in Table V.

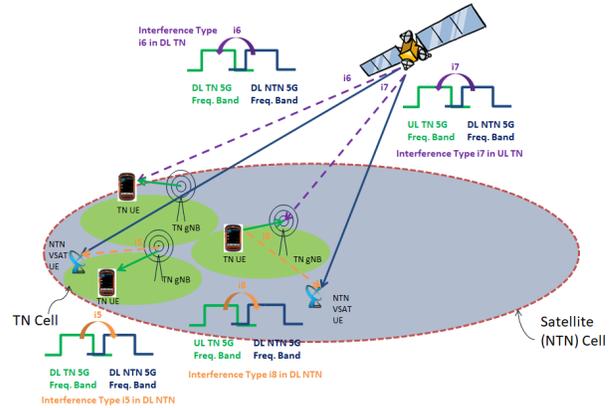

Fig. 4. Exemplification of interference type i5 to i8

TABLE V
TN ACLR AND ACS VALUES USED FOR Q/V BAND

| TN equipment and requirements | | TN (2 GHz) | TN (27 GHz) | 6G NTN assumptions for TN operating at 37, 47 GHz |
|---|---|---|---|---|
| BS | ACLR | 45 dB | 28 dB | 26 dB |
| | ACS | 46 dB | 24 dB | 22 dB |
| UE | ACLR | 30 dB (ACLR1) 43 dB (ACLR2) | 17 dB | 16 dB |
| | ACS | 33 dB | 23 dB | 22 dB |

The coexistence studies for 6G NTN use 3GPP methodology based on NTN service link evaluation (e.g. Coupling Loss, SINR) and 5% throughput loss to derive relevant ACLR and ACS requirements. Moreover, technical reports such as TR 38.811 [18] and TR 38.821 [19] developed as part of Release 15 and Release 16 are relevant references for other NTN parameters and requirements. Interference mitigation solutions such as advanced receivers based on MMSE-IRC [18], improved MAC scheduling, NTN beam management (beam hopping, beam steering), and FFR techniques can also be used to improve coexistence [1].

*B. Parameters to be used for coexistence analysis*

Together with spectrum management scenarios, the simulations developed for 6G NTN shall consider relevant configurations in Q/V bands. For instance, the reference parameters for the NTN terminal in Q/V bands are presented in Table VI, where the VSAT implementation can be fixed or mobile (UE mounted on moving platform). The SAN reference parameters are presented in Table VII for three different satellite orbits (i.e. GEO, LEO@600km, and LEO@1200km).

TABLE VI
Q/V BAND NTN UE PARAMETERS

| Parameters | NTN VSAT |
|---|---|
| Transmit Power | 2 W (33 dBm) |
| Antenna type | 15 cm equivalent aperture diameter (circular polarisation) |
| Antenna gain | $G_{Tx}$ : 35.2 dBi ; $G_{Rx}$ : 32.9 dBi |
| Noise figure | 2 dB |
| Output loss | 1.5 dB |
| EIRP | 36.7 dBW |

TABLE VII
Q/V BAND NTN SAN PARAMETERS

| SAN Parameters | GEO | LEO-1200 km | LEO-600 km |
|---|---|---|---|
| Satellite altitude [km] | 35786 km | 1200 km | 600 km |
| Equivalent satellite antenna aperture [m] | 2.7; 2.1 (DL;UL) | 0.27; 0.21 (DL;UL) | 0.27; 0.21 (DL; UL) |
| Satellite max Tx/Rx Gain [dBi] | 60.4 dBi ($G_{Tx}$; $G_{Rx}$) | 40.4 dBi ($G_{Tx}$; $G_{Rx}$) | 40.4 dBi ($G_{Tx}$; $G_{Rx}$) |
| Satellite EIRP density [dBW/MHz] | 45 | 18 | 15 |
| 3 dB beamwidth [deg] | 0.1884 deg | 1.884 deg | 1.884 deg |
| Satellite beam diameter at nadir [km] | 117.7 km | 39.5 km | 19.7 km |
| Satcom Repeater Noise Figure [dB] | 4 dB | 4 dB | 4 dB |

V. CONCLUSIONS

Emphasizing the full integration of NTN into 6G to support ubiquitous communication and global coverage, this paper provides a profound overview of advancements that are being emerged within the integrated framework of 6G NTN. This involves, at first, the identification of challenges associated with 6G NTN RATs and answers them by advancing the aspects, specifically, RRM, spectrum coexistence in terrestrial and non-terrestrial components, and the efficient waveform design supporting 6G NTN air interface. Based on the extensive discussion provided herein, it is envisaged that the proposed paramount advancements have capabilities to combat the impediments associated to 6G NTN RATs.

ACKNOWLEDGEMENT

This work is performed in 6G-NTN project funded by Smart Networks and Services Joint Undertaking under the EU Horizon Research and Innovation program, Grant No. 101096479, by "Ministerio de Asuntos Económicos y Transformación Digital" and the EU-NextGenerationEU in the frameworks of the PRTP and MRR under reference TSI-063000-2021-70, and by Catalan government SGR-Cat 2021-01207.